%% file: main.tex
\documentclass[aps, reprint, superscriptaddress,prl]{revtex4-2}
\usepackage{siunitx}
\usepackage{graphicx}
\usepackage{placeins}
\usepackage{textcomp}
\usepackage{amssymb}
\usepackage{amsmath}
\usepackage{times, txfonts}
\usepackage{makecell}
\usepackage{physics}
\usepackage[english]{babel}
\usepackage[colorlinks=true,linkcolor=blue,urlcolor=blue,citecolor=blue,pdfusetitle]{hyperref}

\usepackage{braket}
\usepackage{csquotes}
\MakeOuterQuote{"}

\usepackage{amsfonts}
\usepackage{gensymb}
\usepackage{xcolor}
\usepackage{multirow}
\usepackage{mathrsfs}
\usepackage[title]{appendix}
\usepackage{booktabs}
\usepackage{algorithm}
\usepackage{algorithmicx}
\usepackage{algpseudocode}
\usepackage{listings}
\usepackage{cleveref}

\usepackage{amsthm}

\renewcommand{\thefigure}{S\arabic{figure}}

\begin{document}

\title{High-fidelity entangled photon pairs from a quantum-dot-based single-photon source}

\author{Malwina A. Marczak}
\thanks{These authors contributed equally}
\affiliation{Department of Physics, University of Basel, Klingelbergstrasse 82, CH-4056 Basel, Switzerland}
\author{Spencer J. Johnson}
\thanks{These authors contributed equally}
\affiliation{Jet Propulsion Laboratory, California Institute of Technology, Pasadena, 91109, CA, USA}
\affiliation{Department of Physics, University of Illinois Urbana-Champaign, 1101 W. Springfield Avenue, Urbana, 61801, IL, USA}
\author{Mark R. Hogg}
\thanks{These authors contributed equally}
\affiliation{Department of Physics, University of Basel, Klingelbergstrasse 82, CH-4056 Basel, Switzerland}
\author{Timon L. Baltisberger}
\affiliation{Department of Physics, University of Basel, Klingelbergstrasse 82, CH-4056 Basel, Switzerland}
\author{Nathan Arnold}
\affiliation{Department of Physics, University of Illinois Urbana-Champaign, 1101 W. Springfield Avenue, Urbana, 61801, IL, USA}
\author{Benjamin~E.~Nussbaum}
\affiliation{Department of Physics, University of Illinois Urbana-Champaign, 1101 W. Springfield Avenue, Urbana, 61801, IL, USA}
\author{Clotilde M. N. Pillot}
\affiliation{Department of Physics, University of Basel, Klingelbergstrasse 82, CH-4056 Basel, Switzerland}
\author{Sascha R. Valentin}
\affiliation{Lehrstuhl f\"{u}r Angewandte Festk\"{o}rperphysik, Ruhr-Universit\"{a}t Bochum, D-44780 Bochum, Germany}
\author{Arne Ludwig}
\affiliation{Lehrstuhl f\"{u}r Angewandte Festk\"{o}rperphysik, Ruhr-Universit\"{a}t Bochum, D-44780 Bochum, Germany}
\author{Paul G. Kwiat}
\affiliation{Department of Physics, University of Illinois Urbana-Champaign, 1101 W. Springfield Avenue, Urbana, 61801, IL, USA}
\author{Richard J. Warburton}
\affiliation{Department of Physics, University of Basel, Klingelbergstrasse 82, CH-4056 Basel, Switzerland}

\date{\today}

\begin{abstract}
Entangled photon pairs are a ubiquitous resource in quantum technologies, used in quantum key distribution and quantum networking as well as fundamental tests of non-locality. For scalable quantum networks, pairs that are indistinguishable in all unentangled degrees of freedom are essential, as they enable high-fidelity entanglement swapping across network nodes. To date the most-studied sources of “swappable” entangled photon pairs have been based on spontaneous parametric down-conversion (SPDC) in non-linear crystals. However, the probabilistic nature and unavoidable trade-off between brightness and unwanted multi-photon emission limits their performance in lossy channels.
Here, we demonstrate a high-fidelity source of “swappable” entangled photon pairs using a semiconductor quantum dot (QD) coupled to a tunable microcavity.
By actively modulating the QD emission between orthogonal polarisation states, delaying one path in a low-loss Herriott cell, and recombining the two on a balanced beam splitter, we generate entangled photon pairs with a fidelity of 96.1$\pm0.5$\%.
We identify and mitigate fidelity-limiting factors, achieving a maximum fidelity of 98.1$\pm0.5$\% through time-resolved post-selection.
The scheme suppresses residual multi-photon events concentrated near the excitation pulse and has only a modest impact on the rate.
Furthermore, the photons are mutually indistinguishable, enabling efficient entanglement swapping.
Our results establish semiconductor QDs as a viable platform for quantum network-compatible swappable entangled photon pair generation, with feasible entanglement generation rates exceeding 0.5 Gpairs/s.
\end{abstract}

\maketitle

\noindent\textbf{Introduction}\\
Bright, high-fidelity sources of entangled photon-pairs are critical for future quantum networks. Distributed quantum sensing can link telescopes to improve the resolution \cite{Mohageg2022}; nodes connected by secure quantum channels can guarantee the integrity of shared data \cite{Azuma2015, Kolodynski2020}; and network-linked quantum processors can greatly enhance computational power \cite{Main2025}. To date, the most prominent sources of entangled photon pairs employ spontaneous parametric down-conversion (SPDC) \cite{Kwiat1995}. However, the inherent trade-off between brightness and photon number purity limit their utility in mid- to long-range quantum networks. These drawbacks can be addressed by switching to an engineered single emitter to create photons deterministically.
%
% however its probabilistic nature and the possibility of multi-pair events 
%limit their utility in mid- to long-range quantum networks. These drawbacks can be addressed by switching to an engineered single emitter to create photons deterministically. A semiconductor quantum dot (QD) is an attractive choice for the single emitter \cite{Santori2002, Ding2016, Tomm2021}, and can produce entangled photon pairs both directly from a two-photon decay via the biexciton cascade \cite{Akopian2006, Muller2014, Huber2018, Versteegh2014} or from overlapping spectrally unentangled, indistinguishable single photons at a beam-splitter and erasing the which-path information \cite{Fattal2004, Valeri2024}. Although the biexciton cascade has demonstrated high entanglement fidelities \cite{Huber2018}, the two emitted photons are spectrally distinguishable and cannot be freely interfered. 
%
% The second approach produces entangled, indistinguishable photon pairs via a probabilistic process with 50\% success probability. At scale, such sources allow for efficient entanglement swapping, enabling long-range quantum links.
%
% and produces spectrally unentangled, indistinguishable single photons . Entangled photons can be created from the single photons at a beam-splitter , a probabilistic process but with 50\% success probability. At scale, such sources allow for efficient entanglement swapping, enabling long-range quantum links.
A semiconductor quantum dot (QD) is an attractive choice for the emitter, deterministically producing single photons with high purity and indistinguishability \cite{Santori2002, Ding2016, Somaschi2016, Uppu2020, Tomm2021}. QDs can produce entangled photon pairs both directly from a two-photon decay via the biexciton cascade \cite{Akopian2006, Muller2014, Huber2018, Versteegh2014}, or indirectly by interfering spectrally unentangled, indistinguishable single photons on a beam-splitter and erasing any which-path information \cite{Fattal2004, Valeri2024}. Although the biexciton cascade can generate highly entangled photon pairs \cite{Huber2018}, the photons emitted in the two steps of the cascade have different frequencies and thus cannot interfere with each other. Photons from the same transition in separate cascade events can in principle interfere, but their indistinguishability is fundamentally limited by temporal correlations in the emission process \cite{Schoell2020}. Recent work has demonstrated cavity-enhanced spontaneous two-photon emission from the biexciton state, producing spectrally matched entangled pairs \cite{Liu2025}; however, the indistinguishability of these photons has not yet been characterised.

For the interference-based scheme, the entanglement generation process itself relies on two-photon interference between photons independently emitted by the quantum dot. High entanglement quality therefore guarantees that the photons are in pure quantum states, e.g., spectrally unentangled, which is one of the primary conditions for swappability. Realising this approach with a bright source producing highly indistinguishable photons at high rate could thus form a promising route towards scalable quantum networks. Beyond generation rate, entanglement fidelity remains an important factor, as errors accumulate through successive swapping operations and rapidly reduce end-to-end performance. Entanglement swapping with quantum dot photons has been demonstrated using entangled pairs from the biexciton cascade \cite{BassoBasset2019, Zopf2019}, albeit with low-indistinguishability photons, severely limiting both the fidelity and success rate. In contrast, implementations leveraging interference-based schemes remain largely unexplored. Nevertheless, substantial progress has been made in related areas, including probabilistic heralded entanglement generation using quantum-dot-based single-photon sources and the creation of multi-photon entangled states such as Greenberger–Horne–Zeilinger (GHZ) and cluster states \cite{Cao2024, Chen2024, Meng2024, Li2020}. Together, these advances solidify the position of semiconductor quantum dots as one of the most promising platforms for scalable entangled-photon generation and future quantum network applications.

Using a QD-based single-photon source, we overlap successive single photons in orthogonal polarisations at a beam-splitter to create indistinguishable, polarisation-entangled photon pairs, demonstrating post-selected entanglement with a fidelity of $F=96.1 \pm 0.5 \%$. The high efficiency of our source enables the creation of entangled photon pairs at high rates. %significantly higher rates than previously reported; 
We measure a coincidence rate of 320\,kpairs/s; by characterising the post-beam-splitter losses in our system, we infer an entangled pair generation rate of 2.1$\pm0.16$\,Mpairs/s. We present a proof-of-principle experiment to show that this rate can be increased at least 25-fold with no impact on the fidelity by increasing the repetition rate of the excitation laser. We identify the imperfect purity, i.e., $(1-g^{(2)})$ where $g^{(2)}$ is the autocorrelation at zero delay for the QD-based single-photon source, as the main factor limiting the fidelity. Further, we demonstrate two modified post-selection protocols which increase the fidelity up to $98.1 \pm 0.5 \%$ with only a modest impact on the rate (24\% reduction). Finally, we present a detailed quantitative analysis comparing our quantum dot approach to SPDC sources for the network-critical application of entanglement swapping.\\

\begin{figure}
    \centering
    \includegraphics{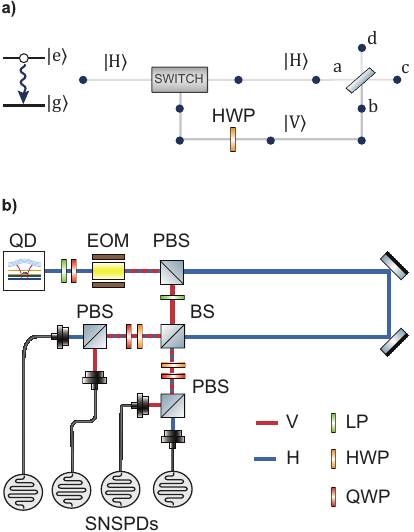}
    \caption{\textbf{Experimental setup.} (a) Schematic of a postselected entanglement source using a stream of indistinguishable photons. (b) Representation of our experimental setup. The polarisation of every other photon from the quantum dot (QD) is switched via the resonant electro-optic modulator (EOM) and the polarising beam splitter (PBS). After recombining at a non-polarising beam splitter (NPBS), the resulting entangled photon pairs are detected via coincidence counting using superconducting nanowire single photon detectors (SNSPDs), while the half-waveplate (HWP), quarter-waveplate (QWP), and PBS in each output arm of the beam splitter are used for quantum state tomography to reconstruct the density matrix.}
    \label{fig:PS_source_QD}
\end{figure}

\noindent\textbf{Entanglement with a QD-based single-photon source}\\
Fig.\,\ref{fig:PS_source_QD}a shows the basic principle of operation to create polarisation entangled photon pairs from a source of indistinguishable single photons. A switch is used to send alternating photons into separate spatial modes. One photon is delayed and its polarisation is rotated, then both photons are interfered on a 50:50 non-polarising beam-splitter (NPBS), resulting in the state 
\begin{equation}\label{eq:ps_state}
    \ket{H}_a\ket{V}_b \rightarrow \frac{1}{2} \big(\ket{H}_c\ket{V}_d + i (\ket{H}_c\ket{V}_c + \ket{H}_d\ket{V}_d) - \ket{H}_d\ket{V}_c\big),
\end{equation}
with mode labels $a$, $b$, $c$, and $d$ defined in Fig.\,\ref{fig:PS_source_QD}a. If a photon is detected at each output port, and if the photons are indistinguishable in all ways except their polarisation, the (normalised) state will be entangled: $\ket{\Psi}^{-} \equiv \frac{1}{\sqrt{2}} \big(\ket{H}_c\ket{V}_d - \ket{H}_d\ket{V}_c\big)$. We note that this is true regardless of the relative phase of the photons before the beam-splitter, and is even independent of the beam-splitter reflectivity (assuming it is independent of polarisation), though the latter will of course affect the post-selection probability. 
%which achieves its maximum value of 50\% for a 50:50 NPBS.

Our source uses an InGaAs QD in an open microcavity \cite{Greuter2015, Tomm2021}, creating single photons at a wavelength of 922.5~nm. The QD is excited resonantly with 5~ps-duration pulses from a mode-locked laser running at a repetition rate of $R_{\rm L} = 76.3$~MHz. The purity $(1-g^{(2)})$ is $98.5 \pm 0.1\%$; the corrected visibility in a Hong-Ou-Mandel interference experiment is $V_{HOM} = 98.1 \pm 1.4$\%, after accounting for finite $g^{(2)}$ and setup imperfections (see Supplementary Information). The end-to-end efficiency $\eta$, the probability of creating a single photon at the output of the final optical fibre following an excitation pulse, is $\eta = 49\pm3$\% in these experiments. We alternate the polarisation and thereby the delay experienced by successive photons using a resonant electro-optic modulator (EOM) and a polarising beam-splitter (PBS), overlapping them pairwise on a 50:50 NPBS (Fig.\,\ref{fig:PS_source_QD}b).

In the data analysis, we retain only the events in which a photon is detected in each output of the beam-splitter, using waveplates and PBSs to perform quantum state tomography and thereby reconstruct the density matrix $\rho$ \cite{James2001}. We use the singlet fraction, defined as the maximum overlap between a state $\rho$ and the closest maximally entangled state \cite{Modlawska2008}, as a metric for the entanglement, as it is robust against any local unitary transformations on the photons, conveying how well $\rho$ could be made to match the singlet state $\ket{\Psi}^{-}$. Below we use the term "fidelity" to describe this overlap.\\

% (b) Pump multiplexing setup used to demonstrate that a repetition rate of a GHz is achievable. A single laser pulse is split into two such that the single-photon source produces two photons separated by 500~ps in time. Subsequently, two closely-spaced photon pairs are created.}
    % \label{scheme}
% \end{figure}

\begin{figure*}[t!]
    \centering
    \includegraphics{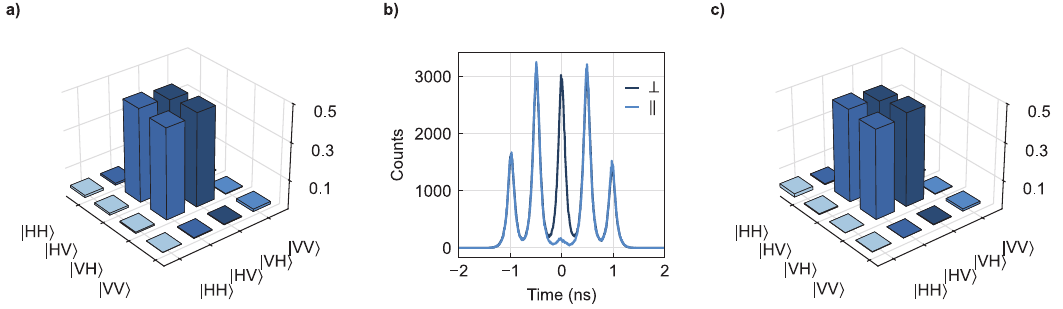}
    \caption{\textbf{Indistinguishable entangled photon pairs.} (a) Two-photon density matrix of the singlet fraction for the source running at repetition rate $R_{L}$, yielding a fidelity of $F=96.1\pm 0.5 \%$ for the state $\ket{\psi} = \ket{HV}+e^{i\phi}\ket{VH}$, where $\phi \approx -11$ degrees (the singlet fraction is obtained by applying single-qubit rotations that optimise the overlap with the target Bell state; this results in a density matrix that has only real components). (b) Hong-Ou-Mandel interference experiment for excitation pulse separation of 500~ps, showing visibility of $V_{HOM} = 97.6 \pm 1.5\%$. (c) Two-photon density matrix of the singlet fraction for excitation pulse separation of 500~ps, with a fidelity of $F = 95.2\pm 0.5 \%$.}
    \label{rho}
\end{figure*}

% \cite{Valeri2024}Not sure where to put this now, since originally it was the citation for the incorrect description of phi.

\noindent\textbf{Results: entanglement creation rate}\\
We run the single-photon source by exciting the QD once per laser pulse, i.e., at rate $R_{\rm L}=76.3$~MHz, such that an attempt is made to create an entangled photon-pair at rate $R_{\rm L}/2$. The reconstructed density matrix shown in Fig.\,\ref{rho}a reveals an entanglement fidelity of $F=96.1 \pm 0.5 \%$ for the state $\ket{\psi} = \ket{HV}+e^{i\phi}\ket{VH}$, where $\phi$ arises from birefringence in or after the beam-splitter. While the tomography setup suffers from losses, these are known such that the detector count rates can be converted into the entangled photon-pair creation rate $R_{\rm E}$ at the beam-splitter output. We find $R_{\rm E} = 2.1\pm0.16$~Mpairs/s, consistent with the 49$\pm$3\% inferred from our independently measured losses before the recombination beam-splitter.
%. Based on a characterisation of the losses before the recombination beam-splitter, an entanglement generation rate of 2.1~Mpairs/s matches with an end-to-end source efficiency of 51\%, consistent with our independently measured value of 49$\pm$3\%.
% Equivalently, the source efficiency along with the modulator efficiency give the same value of $R_{\rm E}.

The most obvious way to increase $R_{\rm E}$ is to increase the efficiency of the source. Using our own best value ($\eta=57$\% \cite{Tomm2021}) with the current setup would increase $R_{\rm E}$ to 2.85~Mpairs/s. This improvement can be combined with a powerful second approach. The radiative lifetime of the QD in the open microcavity is reduced to just 60~ps by the Purcell effect, suggesting that the pump rate $R_{\rm L}$ (determined by the design of the laser and not the QD source) is far too conservative. To explore this concept with the present laser, we split each laser pulse into two and delay one pulse relative to the other such that each laser pulse results in \emph{two} QD excitations, doubling the effective rate of entanglement generation. 
Fig.\,\ref{rho}b shows the Hong-Ou-Mandel interference for a pulse separation of 500 ps achieving $V_{HOM} = 97.6 \pm 1.5 \%$. The reconstructed density matrix for this "doubled" source is shown in Fig.\,\ref{rho}c, with an entanglement fidelity of $F=95.2\pm 0.5 \%$, identical (within error) to the fidelity with a single excitation per pulse. This result validates the concept of exciting the QD at $R_{\rm L}$ = 2 GHz. With $\eta=49/57/71.2$\% (the latter being the current record quantum dot source efficiency \cite{Ding2025}), $R_{\rm E}$ then increases to 55/74.6/116.4~Mpairs/s. Improving optical component losses in the system (see Supplementary Table 1), we could improve the setup efficiency by up to 40\%, leading to entanglement generation rates up to 215\,Mpair/s. Finally, such improvements can be further extended to {\it two} quantum dots, one emitting H-polarised and one V-polarised (but otherwise mutually indistinguishable \cite{Zhai2022}) photons, thus obviating the  need for an active optical switch to alternate the polarisation of successive photons. This configuration would enable a feasible swappable entangled-pair generation rate of over 0.5~Gpairs/s, far beyond what any SPDC entangled source has demonstrated.\\

\begin{figure}[t!]
    \centering
    \includegraphics{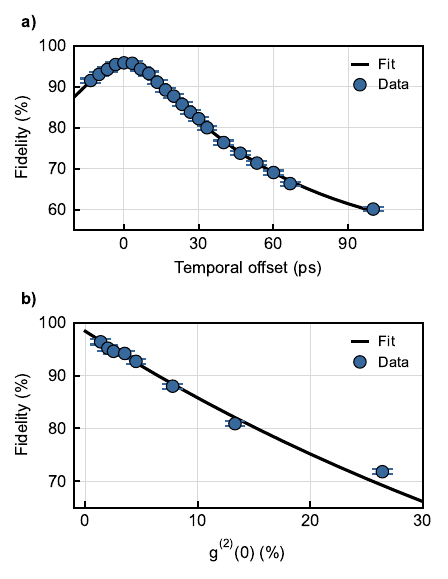}
    \caption{\textbf{Dependence of fidelity on source properties.} (a) Fidelity as a function of the temporal misalignment between the two interfering photons. At zero delay ($\Delta t = 0$ ps), the fidelity is $F = 95.8\pm 0.5 \%$. It decreases to $F = 69.1\pm 0.5 \%$ at $\Delta t = 60$ ps, corresponding to one radiative lifetime. (b) Fidelity versus $g^{(2)}(0)$. For $g^{(2)} = 1.3\%$, the fidelity is $96.4\pm 0.5 \%$, while for $g^{(2)} = 7.7\%$, $F$ decreases to $88.8\pm 0.5 \%$. In both panels, black lines represent theoretical fits (see Supplementary Information).}
    \label{homg2}
\end{figure}

\noindent\textbf{Results: entanglement fidelity}\\
We turn now to the entanglement fidelity, $F$, and the main mechanism limiting its value.  First, we measure $F$ as a function of the temporal offset between the interfering photons by introducing a temporal misalignment of the interferometer paths, Fig.\,\ref{homg2}a. As expected, the fidelity decreases as the temporal offset increases, exhibiting the maximum when the photons arrive simultaneously at the beam-splitter. This behaviour is consistent with the finite temporal wave packet of the photons, the offset reducing their indistinguishability and therefore the observed entanglement fidelity. Second, we measure the entanglement fidelity versus $g^{(2)}(0)$ values, see Fig.\,\ref{homg2}b. We use an excitation pulse area below $\pi$ to create $g^{(2)}(0)$ values below $2\%$. For $g^{(2)}(0)$ values above $2\%$, we gradually increase the leakage of the excitation laser, observing that the fidelity decreases as $g^{(2)}$ increases. The decrease in fidelity at high $g^{(2)}(0)$ values  highlights the scheme's vulnerability to unwanted multi-photon events, either from background or multiple QD excitations. 

\begin{figure}[t!]
    \centering
    \includegraphics{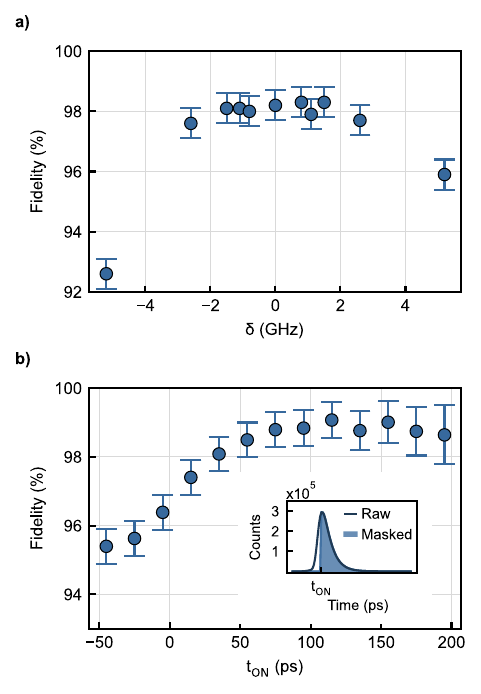}
    \caption{\textbf{Improving fidelity with filtering.} (a) Entanglement fidelity as a function of etalon detuning for the 5-GHz bandwidth etalon. Zero detuning corresponds to the etalon being resonant with the single photons. The maximum fidelity is $98.3\%$ for $\delta = 0.8$~GHz. The asymmetry in the entanglement fidelity results from the asymmetric transmission of the etalon. (b) Main plot: two-photon entanglement fidelity versus switching time, $t_{\rm ON}$. In the data analysis, coincidences for $t<t_{\rm ON}$ are rejected and coincidences for $t \ge t_{\rm ON}$ are retained. $t=0$ represents the time of the maximum intensity of the laser pulse (see Supplementary Information). Inset: Example histogram of the timing between the data and trigger signals. The dark blue curve corresponds to the raw histogram, while the lighter blue curve shows the histogram after temporal filtering. Only counts within the shaded region are included in the fidelity calculation; counts in the unshaded region are excluded.}
    \label{switch}
\end{figure}

The spurious two-photon events have a negative impact on the entanglement fidelity as they result in an extra photon in one output port of the beam-splitter, adding, e.g., terms like $\ket{HH}$ or $\ket{VV}$ to the output state. Eliminating the two-photon events, i.e. decreasing $g^{(2)}$, therefore results in an increase in the entanglement fidelity. As a diagnostic tool, we insert an etalon between the single-photon source and the modulator. The etalon has a bandwidth of 5~GHz, comparable to the 3-GHz bandwidth of the QD photons, and is tuned into resonance with the QD source. We find that with the etalon, $g^{(2)}$ reduces from 2.0\% to just 0.9\%, and $F$ increases from $96.1 \pm 0.5\%$ to $98.3 \pm 0.5\%$, Fig.\,\ref{switch}a. This solution is not a practical way forward as the etalon we used reduced the entanglement rate $R_{\rm E}$ significantly. However, the results provide insight. The reduction in $g^{(2)}$ shows that the etalon eliminates most of the two-photon events in the source's output. In turn, this result tells us that at least one of the photons in a two-photon event has a large bandwidth such that it is transmitted through the etalon with low probability. 

These observations are consistent with a model of the QD source: the two-photon events result from a double excitation whereby one laser pulse creates two photons \cite{Gonzalez-Ruiz2025}. The first photon has to be created in a very short time in order for the QD to return to its ground state and be re-excited within the duration of a single laser pulse. Hence, the first photon has a large bandwidth with respect to the source's single photons, and is therefore rejected by a suitably chosen spectral filter.

This picture suggests an alternative strategy to improve the entanglement fidelity. Specifically, time gating can be used to exclude coincidences from these large-bandwidth "early" photons, which occur within $\sim$5 ps of the excitation pulse (compared to single-photon events, which on average occur $\sim$60 ps after the pulse) \cite{Walther_temporal_filtering_2025}. This strategy relies on post selection, but given that the entanglement scheme itself relies on post selection no particular drawback is introduced. The only proviso is that the timing jitter of the detector hardware is smaller than the radiative lifetime, here 60~ps. 
We implemented this idea by running the experiment in time-tag mode to record the exact time of each detection event. Subsequently, we reconstruct the density matrix and the entanglement fidelity by passing the data through an off-on-off temporal filter. The timing of the off-on-off switch, $t_{\rm ON}$, is varied with respect to time zero (see Supplementary Information). Fig.\,\ref{switch}b shows the entanglement fidelity versus $t_{\rm ON}$. In practice, the timing jitter of the hardware, 35~ps, broadens the temporal filter around $t_{\rm ON}$.  We find that the fidelity increases as the switching time is changed from $-45$ to $+35$~ps with only modest increases at larger values. This behaviour is consistent with the notion that the large bandwidth photon is created at very small times, supporting the reexcitation model. From the applications point of view, the scheme has the advantage that the fidelity is boosted without introducing any additional loss in the hardware, unlike the approach with the etalon for instance. In practice, we find that modified post-selection procedure results in an increase in entanglement fidelity from $95.5 \pm 0.5\%$ to $98.1 \pm 0.5\%$, while only modestly decreasing $R_{\rm E}$, by 24\%.\\

\noindent\textbf{Comparison between QD and SPDC sources}\\
%REWRITE THIS: The maximum creation rate of entangled photon pairs in this scheme is $\frac{1}{2}\eta^2$. In the best case, $\eta=57\%$, i.e., the efficiency immediately after the source. We consider implementing the protocol via a lossy channel, e.g., a long optical fibre, such that $\eta$ is much reduced. We now compare the rates to those achieved with an SPDC source. If single-photon detectors without photon-number resolving capability, as is usually the case, the rate with the QD source exceeds that of an SPDC source for all losses. However, if single-photon detectors with photon-number resolving capability are employed, the SPDC rate can be boosted. The reason is that in a heralding scheme, the creation of multi-photon events can be detected and discarded, allowing the probability of a single-photon events to be increased without cost. Once a lossy link is included, this potential advantage is lost and a QD source out-performs a SPDC source.

The maximum pair rate for a single entanglement source based on this scheme is $\frac{R}{2}\eta_{s}^2$, where $\eta_{s}$ is the collection efficiency of each photon and $R$ is the repetition rate of the pump laser. Considering an entanglement source based on SPDC, the maximum pair rate in the low-pump-power regime is approximately $RP_1\eta_{s}^2$, where $P_1$ is the probability of producing a single pair in each pulse, typically $1-5\%$ \cite{Dhara2022} depending on the application. However, SPDC-based sources benefit greatly from the implementation of photon-number-resolving (PNR) detectors, allowing events in which multiple pairs are produced within a single pulse to be discarded. With high source and detector efficiencies, single-pair probabilities in excess of $10-20\%$ are achievable while keeping multi-pair errors low. Combined with multiplexing techniques, near unity-efficiency single-photon sources are possible \cite{Kaneda2019}, though have not been demonstrated experimentally with entangled states. 
%(IS THIS TRUE? I (PGK) THINK SO...) 

\begin{figure}[t!]
    \centering
    \includegraphics[width=1\linewidth]{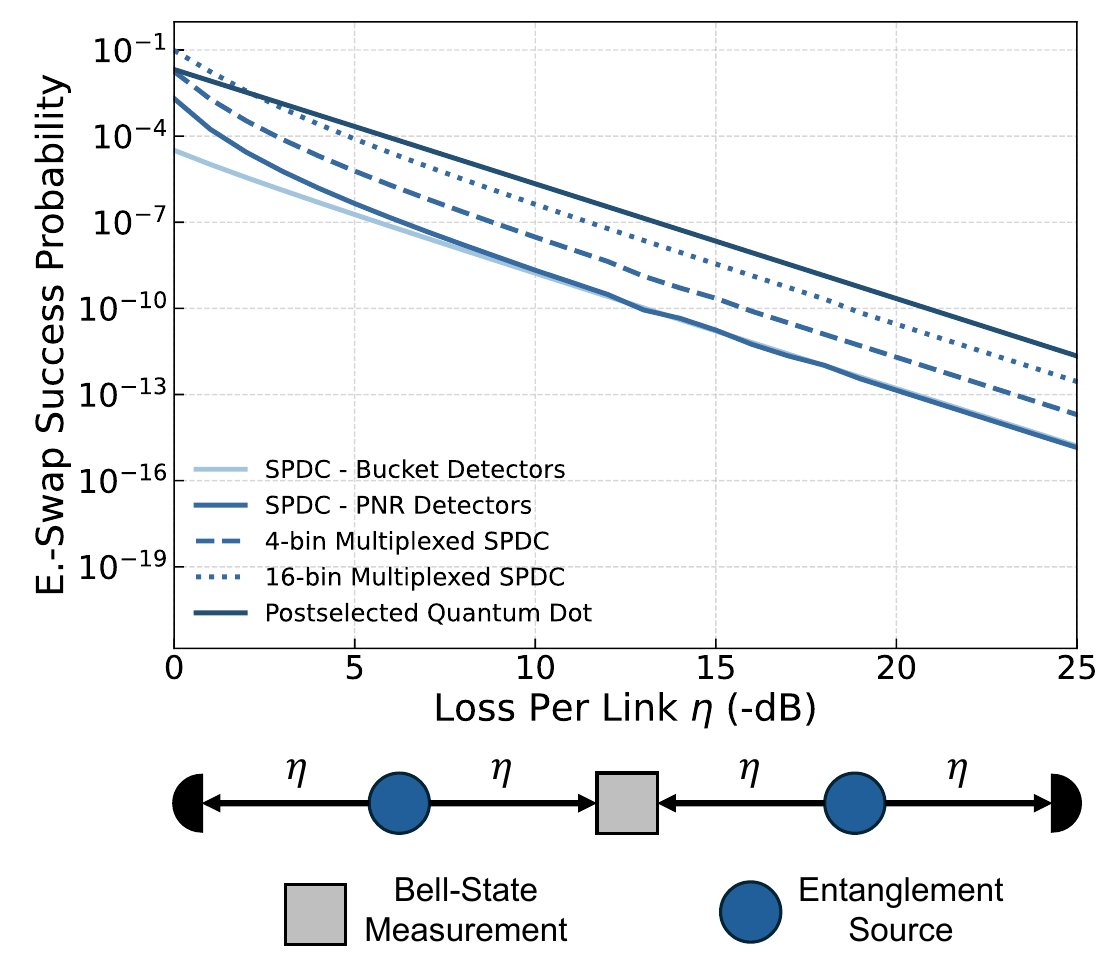}
    \caption{\textbf{Entanglement-swapping rates across platform.} Entanglement-swapping (E.-Swap) rate comparison between a postselected quantum-dot source and various SPDC-based sources. \emph{Plot assumptions:} SPDC extraction efficiency of 0.8; optimised SPDC pump power to achieve entanglement fidelity $F > 97\%$; QD extraction efficiency of 0.71; spatial multiplexing with insertion efficiency of 1 and switch efficiency of 0.97; detector efficiency of 0.9 with photon-number resolution (PNR).}
    \label{swapping}
\end{figure}

The real benefit of the quantum dot source over SPDC comes when we look at any application requiring the interference of photons from two or more sources. Consider the application of entanglement swapping, where photons from two separate entangled sources are mixed on a beam-splitter. A successful Bell state analysis (which can be successful $50\%$ of the time) indicates that the two "outer" photons (to which the ones just measured after the beam-splitter are entangled) are now entangled \cite{BSM_1998}; see Fig.\,\ref{swapping} sketch. The problem with SPDC sources is that there is essentially as high a likelihood for the outcome where one of the sources produced two pairs of photons, as there is for the desired outcome in which each source produced a single pair. These double-pair events dramatically reduce the fidelity of the swapped entanglement, particularly in lossy environments. Fig.\,\ref{swapping} shows an entanglement swapping comparison between SPDC (including various levels of multiplexing), and the quantum dot source methods described here; we see that the latter are beneficial -- yielding up to a two-order of magnitude final rate enhancement -- unless the SPDC sources incorporate large amounts of very efficient multiplexing. With the assumption that both sources produce completely indistinguishable photons, the fidelity of the final swapped state is $F_{QD} \approx 97\%$, with the pair probability of the SPDC-based sources optimized to maximize swap rate while keeping the final fidelity $F_{SPDC} \ge F_{QD}$. \\

\noindent\textbf{Conclusions and outlook}\\
We report the creation of post-selected entangled photon-pairs from a quantum-dot-based single-photon source. The high efficiency and repetition rate of the source allows us to create entangled pairs at a rate of 2.1$\pm0.16$~Mpairs/s. %We present a method to increase this rate by a factor of up to 65, by increasing the repetition rate of the source, and demonstrate its feasibility up to 25x enhancement. 
We present a method to increase this rate by increasing the repetition rate of the source, which we demonstrate as a proof-of-concept yielding a 25-fold enhancement. By additionally combining this with the highest reported source efficiency, the rate can be increased by up to 65-fold.
We also outline a scheme based on two separate single-photon sources, which could push the generation rate to the pump repetition rate. We show that the entanglement fidelity is limited largely by two-photon events in the source's output, i.e., by a non-ideal purity $(1-g^{(2)})$. Furthermore, we show that one of the photons in each (unwanted) two-photon event is created in a small time-window around the trigger and has a large spectral bandwidth. By adapting the post-selection protocol to exclude coincidences at very early times in each cycle, we increase the fidelity from $95.5 \pm 0.5\%$ to $98.1 \pm 0.5\%$ at a modest cost to the entanglement creation rate. Alternatively, applying a spectral filter achieves $98.3 \pm 0.5 \%$. Overall, our results show that a high-efficiency, low-noise quantum-dot-based single-photon source can be used to create entangled photons that are suitable for entanglement swapping. Moreover such a source is superior for this application to SPDC entangled photons, unless a large amount of multiplexing is used to boost the SPDC performance.\\

\noindent\textbf{Methods}\\
\textbf{Experimental set-up}
\\
The single-photon source is triggered with a mode-locked Ti:sapphire laser operating at repetition rate $R_L=76.3$~MHz. Photons are detected with four SNSPDs (superconducting-nanowire single-photon detectors) with efficiencies around 82\%. The combined timing jitter of the SNSPDs and photon-counting hardware is 35~ps (see Supplementary Information, Fig.S5). 
The photons emerge from the source with H polarisation and in a single spatial mode. The polarisation of every other photon is switched to V by a bulk resonant electro-optic modulator (QUBIG AM-925SP), with -0.04~dB insertion loss and 200:1 polarisation extinction ratio. To apply the alternate switching precisely over long periods of time, the modulator is frequency-locked to the mode-locked laser. The H photons are delayed by exactly $\Delta T = 1/R_L$ relative to the V photons. The H-polarised and the V-polarised photons enter the two inputs of a "50:50" non-polarising beam-splitter: we measured $R_H = 50.8\pm1\%$; $T_H = 49.2\pm1\%$; $R_V = 49.9\pm1\%$; $T_V = 50.1\pm1\%$. See Supplementary Information for full details. 
The crucial alignment step is to ensure excellent temporal and spatial  overlap of the H- and V-polarised photons at the beam-splitter. The temporal overlap is ensured by recording decay curves on the H- and V-polarised photons separately, adjusting the delay until the decay curves match, resulting in a temporal mismatch $<$3~ps. Spatial overlap is initially achieved by coupling laser light into the output of two of the detection fibres, thereby sending the light "backwards" to the output fibre of the single-photon source, a process which relies on the single-mode characteristic of the fibres. Final alignment is performed by optimising the classical interference visibility, measured with backward-propagating laser light at the main beam splitter. The tomography setup, shown in (Fig.\,\ref{fig:PS_source_QD}b), allows coincidences to be recorded in all basis states \cite{James2001}. The density matrix is reconstructed from the coincidences and from it the singlet fraction is calculated.
Uncertainties are estimated from the standard deviation of repeated measurements.
\\
\textbf{Spectral and temporal filtering}
\\
For spectral filtering, an etalon is inserted as a diagnostic tool between the output of the single-mode output fibre and the electro-optic modulator. The etalon has a free spectral range of 100~GHz, a full-width-at-half-maximum linewidth of 5.2~GHz, (for comparison, the single photons have a FWHM spectral bandwidth of 3~GHz) and on-resonance transmission of $30\%$ . The etalon is temperature-tuned into resonance with the single-photon source; in practice, the free spectral range can be covered in a temperature range from 15 to 55~$^{\rm o}$C.
\\
Temporal filtering is applied in post-processing to the raw time-tag data. A rectangular inclusion window $[t_{ON},\, t_{OFF}]$ is defined relative to the excitation pulse arrival time, and only detection events falling within this window are retained for coincidence analysis and state tomography. This procedure selectively removes early-time multi-photon events concentrated near the excitation pulse, improving entanglement fidelity at the cost of a modest reduction in detection rate (Supplementary Information, Fig.~S6).

\noindent\textbf{Acknowledgements}\\
The work in Basel was funded by Swiss National Science Foundation project 200020\_204069 and innosuisse project SparQ (104.905 IP\_ENG). S.R.V. and A.L. acknowledge support from DFH/UFA CDFA05-06, DFG TRR160, DFG project 383065199 and BMBF-QR.X Project 16KISQ009. S.J.J., N. A., B.E.N., and P.G.K. were supported by Q-NEXT, a U.S. Department of Energy Office of Science National Quantum Information Science Research Center under Award Number DE-FOA-0002253; and by Air Force/Space Force Research STTR Phase II grant SF22D-T004 to Icarus Quantum, Inc. Part of this research was carried out at the Jet Propulsion Laboratory, California Institute of Technology, under a contract with the National Aeronautics and Space Administration (80NM0018D0004).

\bibliography{references}

\newpage
\input{supplementary_information}

\end{document}

%% file: supplementary_information.tex
\setcounter{figure}{0}
\renewcommand{\thefigure}{S\arabic{figure}}
\onecolumngrid
\onecolumngrid
\begin{center}
{\large \textbf{Supplemental Material: High-fidelity entangled photon pairs 
from a quantum-dot-based single-photon source}}\\[0.8em]
{\normalsize 
Malwina A. Marczak,$^{1,*}$ 
Spencer J. Johnson,$^{2,3,*}$ 
Mark R. Hogg,$^{1,*}$ 
Timon L. Baltisberger,$^{1}$ 
Nathan Arnold,$^{3}$ 
Benjamin E. Nussbaum,$^{3}$ 
Clotilde M. N. Pillot,$^{1}$ 
Sascha R. Valentin,$^{4}$ 
Arne Ludwig,$^{4}$ 
Paul G. Kwiat,$^{3}$ 
and Richard J. Warburton$^{1}$
}\\[0.5em]
{\small
$^{1}$Department of Physics, University of Basel, Klingelbergstrasse 82, CH-4056 Basel, Switzerland\\
$^{2}$Jet Propulsion Laboratory, California Institute of Technology, Pasadena, 91109, CA, USA\\
$^{3}$Department of Physics, University of Illinois Urbana-Champaign, 1101 W. Springfield Avenue, Urbana, 61801, IL, USA\\
$^{4}$Lehrstuhl f\"{u}r Angewandte Festk\"{o}rperphysik, Ruhr-Universit\"{a}t Bochum, D-44780 Bochum, Germany\\[0.3em]
}
\end{center}
\vspace{1em}
\section{Experimental set-up}

The single-photon source (SPS) is triggered with a mode-locked Ti:sapphire laser (Coherent Mira 900) operating at a repetition rate of $R_L = 76.3$~MHz. The SPS consists of a semiconductor bottom mirror with InGaAs QDs embedded in an n-i-p diode structure and a dielectric top mirror, as described in \cite{Tomm2021}. During operation, the cavity length is adjusted such that one of the cavity modes is tuned into resonance with the positive trion $(X^+)$ transition of a QD emitting near $\lambda \approx 923$~nm. The cavity supports two orthogonally polarised modes separated by $\sim 50$~GHz; we use the “blue-collection’’ configuration, in which the higher-energy mode is resonant with the QD transition while excitation is performed through the lower-energy mode \cite{Javadi2023}. Laser background is strongly suppressed using a cross-polarised dark-field microscope arrangement \cite{Kuhlmann2013}, yielding a signal-to-background ratio of $\sim350:1$ (verified by turning off the QD emission via voltage tuning) under $\pi$-pulse excitation.

\begin{figure}[t]
    \centering
    \includegraphics[width=0.9\linewidth]{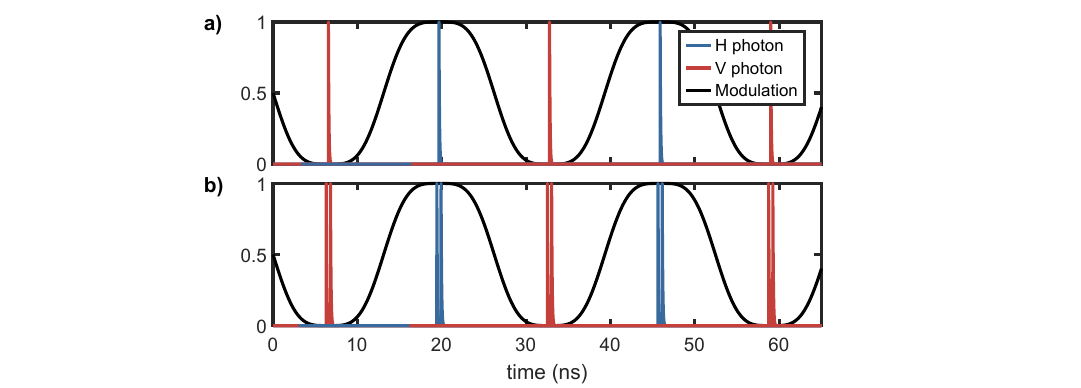}
    \caption{Polarisation modulation timing diagram. (a) Modulation for standard laser repetition rate $R_{\rm L}=76.3$~MHz. Consecutive photons are switched between H (blue curve) and V (orange curve). The black curve shows the projection of the modulator output state onto H as a function of time. (b) Pulse-doubled modulation: here two photons of the same polarisation are output for each modulator switching cycle.
    }
    \label{fig:photon_switching}
\end{figure}

The emitted photons are coupled into single-mode fibre (Thorlabs 780HP) for delivery to the state-preparation stage. A schematic of the experimental setup is shown in Fig.~1 of the main text. A linear polariser (LP) and a quarter-wave plate (QWP) are used to create circularly polarised photons as input to a resonant electro-optic modulator modulator (custom Qubig polarisation modulator, Qubig AM-925SP). The modulator acts as a fast, switchable quarter-wave plate (QWP), producing a string of alternating H- and V-polarised output photons. Figure\,\ref{fig:photon_switching} shows a schematic of the modulation process, with Fig.\,\ref{fig:photon_switching}\,(a) showing the switching for our standard laser repetition rate $R_{L}=76.3$~MHz and Fig.\,\ref{fig:photon_switching}\,(b) for the pulse-doubled case with a 500-ps pulse separation. The modulator electronics are synchronised with the laser clock, and the driving phase and amplitude are adjusted to ensure a polarisation extinction ratio on the order of 200:1. For the highest-rate experiments, the LP and QWP are removed and a manual fibre-based polarisation controller (Thorlabs FPC) is used to create circularly polarised input photons.

Consecutive photons are separated on a polarising beam splitter (PBS) (Thorlabs PBS252), where the horizontally polarised photons are directed into the long arm of the interferometer. The temporal delay ($13.1$~ns) in the long arm is implemented using a compact Herriott cell \cite{Robert_Herriot_cell,Arnold2023}. The cell consists of two opposing 2'' spherical mirrors that produce a sequence of reflections of the propagating beam. This geometry effectively “folds’’ the optical path, allowing us to achieve a long propagation distance without requiring a physically long laboratory setup. Unlike the traditional Herriott configuration—where the beam enters and exits through the same mirror—we employ a scheme in which the photons enter through an aperture in the first mirror and exit through a corresponding aperture in the second mirror. This configuration simplifies the optical alignment and ensures low-loss transmission while still providing the required temporal delay. Two lenses ($f=300$~mm and $f=175$~mm) placed in the long arm compensate for the finite beam divergence, ensuring that the beam waist at the recombining beam splitter closely matches that of the short arm for optimal spatial mode overlap. Although the PBS provides a high extinction ratio ($1000:1$) for the transmitted port, its extinction for the reflected port is comparatively lower ($40:1$). To ensure well-defined polarisation states in both arms, we insert an additional linear polariser (Thorlabs LPNIR100) in the short arm, which improves the overall extinction ratio and balances the polarisation purity of the two paths.

The two consecutively emitted photons are then recombined and made to interfere on a non-polarising beam splitter (Thorlabs CM1-BS014/M), $R_H = 50.8\pm1\%$; $T_H = 49.2\pm1\%$; $R_V = 49.9\pm1\%$; $T_V = 50.1\pm1\%$. Each of the two output ports of the BS is directed into a polarisation-tomography module consisting of a half-wave plate (Thorlabs AHWP10M-980) and a quarter-wave plate (Thorlabs AQWP10M-980) mounted in motorised rotation mounts (Thorlabs K10CR1/M), followed by a PBS (Thorlabs PBS252). After the PBS, the photons are coupled into single-mode fibres (Thorlabs 780HP) and guided to the superconducting nanowire single-photon detectors (Single Quantum EOS). Detector count rates and coincidence rates are recorded using a time-tagging system (Swabian Time Tagger Ultra Performance) connected to the SNSPD driver outputs.

\pagebreak 

The measured efficiency of each element in the system is given in Table~\ref{tab:loss_budget}. For an excitation laser repetition rate $R_{\rm L}$, the entangled photon generation rate at the output of the recombination beam-splitter is given by $(R_{\rm L}/4)\cdot\eta_{sps}^2\cdot\eta_{switch}^2\cdot\eta_{long}\cdot\eta_{short}\cdot\eta_{BS}^2$, where $\eta_{sps}$ is the single photon source end-to-end efficiency, $\eta_{switch}$ is the efficiency of the optics required for polarisation switching, $\eta_{long}$ is the transmission efficiency of photon propagating through the delayed path and $\eta_{short}$ the non-delayed path. For our measured efficiencies, we estimate an entangled photon pair generation rate at the beam-splitter output of $2.1\pm0.26$\,Mpair/s for $R_{\rm L}=76.3$\,MHz. We can also estimate the entangled pair generation rate at the beam-splitter from the measured coincidence count rates; for our measured rate of 320\,kpair/s this ``back-propagation'' method gives $2.1\pm0.16$\,Mpair/s, in excellent agreement with the ``forward-propagation'' estimate.

\begin{table}[htbp]
    \centering
    \caption{Loss budget}
    \label{tab:loss_budget}
    \setlength{\tabcolsep}{18pt}
    \renewcommand{\arraystretch}{1.5}
    \begin{tabular}{lcc}
        % \toprule
         & & \textbf{Efficiency}  \\
        \midrule
        \textbf{Single photon source} &   &  0.49\,$\pm$\,0.03   \\
        \midrule
        \textbf{Components} & Fiber mating sleeve + polarisation paddles  &  0.83\,$\pm$\,0.02   \\
        & Lens pair                                   & 0.97\,$\pm$\,0.002   \\
        & EOM                                         & 0.997\,$\pm$\,0.002  \\
        & PBS                                         & 0.988\,$\pm$\,0.002  \\
        & Delayed interferometer arm                  & 0.913\,$\pm$\,0.005   \\
        & Non-delayed interferometer arm              & 0.987\,$\pm$\,0.005    \\
        & Recombination beam-splitter                 &  0.90\,$\pm$\,0.01 \\
        & Tomography HWP + QWP + PBS                  &  0.90\,$\pm$\,0.01 \\
        % \midrule
        & Fiber coupling + fiber transmission $\rightarrow$ Detector 1         &  0.50\,$\pm$\,0.02 \\
        & Fiber coupling + fiber transmission $\rightarrow$ Detector 2         &  0.536\,$\pm$\,0.02 \\
        % & Fiber coupling (from non-delayed arm) + fiber transmission $\rightarrow$ Detector 1         &  0.445 \\
        % & Fiber coupling (from non-delayed arm) + fiber transmission $\rightarrow$ Detector 2         &  0.442 \\
        \midrule
        \textbf{Detectors} & Detector 1 efficiency                       & 0.9\,$\pm$\,0.03 \\
        & Detector 2 efficiency                       & 0.78\,$\pm$\,0.03 \\
        \bottomrule
    \end{tabular}
\end{table}

\section{Photon purity and indistinguishability}
\begin{figure}[t]
    \centering
    \includegraphics{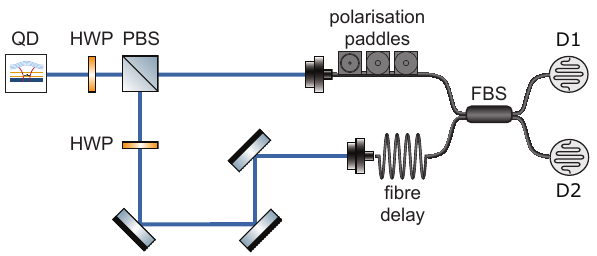}
    \caption{Schematic of the Mach-Zehnder interferometer for Hong-Ou-Mandel measurements. Photons emitted by the quantum dot (QD) are split at a polarising beam splitter (PBS) after passing through a half-wave plate (HWP). An additional HWP in the long arm enables polarisation switching. Both paths are coupled into single-mode fibres, with the short arm containing polarisation paddles for polarisation adjustment and the long arm incorporating a fibre delay. The adjustable path delay enables measurements at different temporal separations ($13.1$~ns for standard repetition rate, 500~ps for pulse-doubling experiments). Photons recombine at a fibre beam splitter (FBS) and are detected by single-photon detectors $D_1$ and $D_2$.}
    \label{fig:hom_setup}
\end{figure}

Photon indistinguishability is quantified via Hong-Ou-Mandel (HOM) interference. Photons are out-coupled from a single-mode fibre and directed into the Mach-Zehnder interferometer shown in Fig.~\ref{fig:hom_setup}. The adjustable path length in the long arm sets the temporal delay between interfering photons. A HWP, PBS and polarisation paddles are used to maximise the classical interference visibility $(1-\epsilon)$. An additional HWP in the long arm switches between co-polarised and cross-polarised configurations.

For standard repetition rate measurements, the path length difference matches the laser repetition period \mbox{($\sim13.1$~ns)}, such that photons from consecutive pulses interfere at the beam splitter. For pulse-doubling experiments, each excitation pulse generates a pair of time-bin encoded photons separated by 500~ps (early and late photon). The interferometer path length is set such that the late photon from the short arm (late-short) interferes with the early photon from the long arm (early-long), with the remaining combinations %(early-short with late-long, and early/late-short with early/late-long) 
(early-short and late-long) being temporally distinguishable and not contributing to the HOM dip. 

For HOM measurements, coincidence counts are recorded for both co-polarised and cross-polarised configurations. To extract the raw visibility, each coincidence peak is integrated over a window of one laser repetition period centred on the peak. The area of each peak is then normalised by the mean area of the surrounding side peaks. The raw visibility is obtained as $V_{  raw} = 1 - A_{\parallel}/A_{\perp}$, where $A_{\parallel}$ and $A_{\perp}$ are the normalised central-peak areas in the co-polarised and cross-polarised configurations, respectively. The corrected HOM visibility $V_\mathrm{HOM}$ accounts for finite $g^{(2)}(0)$, classical visibility $(1-\epsilon)$, and imperfect beam splitter reflectivity ($R$) and transmissivity ($T$) following Ref.~\cite{Tomm2021}, Eq.~(\ref{eq:corr_hom}):

\begin{equation}
    V_\mathrm{HOM} = \frac{1}{(1-\epsilon)^2} \left( \frac{R^2+T^2}{2RT} \right) [1+2g^{(2)}(0)] V_\mathrm{raw}
    \label{eq:corr_hom}.
\end{equation}

\begin{figure}[h]
    \centering
    \includegraphics[scale=0.85]{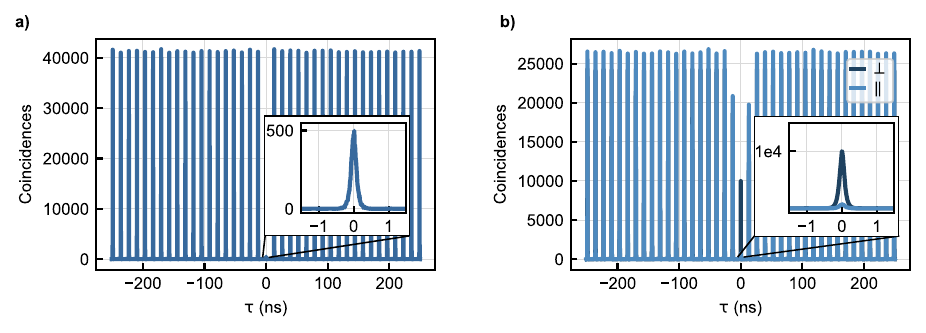}
    \caption{Photon statistics and indistinguishability at standard repetition rate (76.3~MHz). (a) Second-order correlation $g^{(2)}(\tau)$ demonstrating single-photon purity with $g^{(2)}(0) = 1.5 \pm 0.1\%$ (inset: central peak). (b) Hong-Ou-Mandel interference showing two-photon quantum interference (inset: central peak). The raw visibility $V_{  raw} = 91.5 \pm 0.9\%$ is corrected for classical visibility $(1-\epsilon) = 98.5 \pm 0.5 \%$, beam splitter asymmetry ($R = 46.7 \pm 0.9 \%$, $T = 53.3 \pm 0.9\%$), and finite $g^{(2)}(0) = 1.5 \pm 0.1\%$, yielding $V_\mathrm{HOM} = 98.1 \pm 1.4\%$.}
    \label{fig:hom_g2_13}
\end{figure}

\begin{figure}[h!]
    \centering
    \includegraphics[scale=0.85]{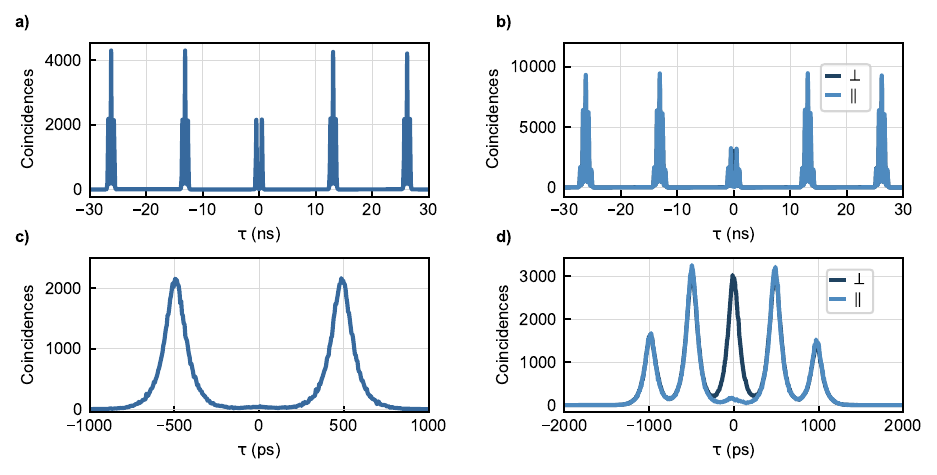}
    \caption{Photon statistics and indistinguishability for pulse-doubling experiments (2.00~GHz repetition rate). (a) Second-order correlation $g^{(2)}(\tau)$ demonstrating single-photon character with $g^{(2)}(0) = 1.2 \pm 0.2\%$. (b) Hong-Ou-Mandel interference showing two-photon quantum interference. The raw visibility $V_{  raw} = 91.2 \pm 0.9\%$ is corrected for classical visibility $(1-\epsilon) = 98.3 \pm 0.5\%$, beam splitter asymmetry ($R = 46.7 \pm 0.9\%$, $T = 53.3 \pm 0.9\%$), and finite $g^{(2)}(0) = 1.2 \pm 0.2\%$, yielding $V_\mathrm{HOM} = 97.6 \pm 1.5\%$. (c), (d) Zoom ins into the central regions of (a), (b).}
    \label{fig:hom_g2_500}
\end{figure}

To measure $g^{(2)}(0)$ of the single-photon source, one interferometer arm is blocked and the temporal correlation $g^{(2)}(\tau)$ between detectors $D_1$ and $D_2$ is recorded. The photon purity is extracted by integrating each correlation peak over a window matching the pump laser repetition period, normalising by the mean area of the side peaks, and taking the normalised central peak area at $\tau = 0$ as $g^{(2)}(0)$. Figure~\ref{fig:hom_g2_13} presents the photon purity and indistinguishability measurements at standard repetition rate. For direct comparison of the co- and cross-polarised HOM measurements, which were recorded at different count rates, we rescale the cross-polarised coincidence histogram by a normalisation factor obtained from the ratio of the integrated areas of the first side peaks in the two measurements. The area of the first side peak in the co-polarised case is 342394.5 counts, while for cross-polarised it is 5940356.0 counts, which gives the normalisation factor of 17.35. The measured $g^{(2)}(0) = 1.5 \pm 0.1\%$ confirms weak multi-photon emission, while the corrected HOM visibility $V_\mathrm{HOM} = 98.1 \pm 1.4\%$ demonstrates high two-photon indistinguishability. 
We repeat the same rescaling procedure for the pulse-doubling experiments at 2.00-GHz repetition rate (Fig.~\ref{fig:hom_g2_500}). The area of the first side peak in the co-polarised case is 111802.5 counts, while for cross-polarised it is 199306.0 counts, which gives the normalisation factor of 1.78. We obtain $g^{(2)}(0) = 1.2 \pm 0.2\%$ and $V_\mathrm{HOM} = 97.6 \pm 1.5\%$, showing that photon quality is maintained at higher generation rates.

\pagebreak

\section{Timing jitter characterisation}
\begin{figure}[t]
    \centering
    \includegraphics{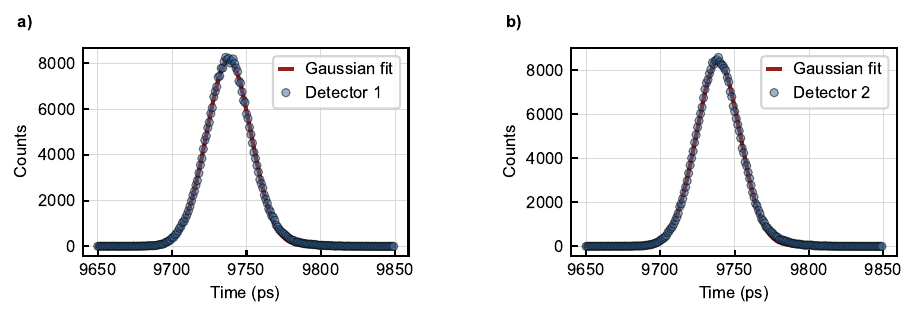}
    \caption{Instrument-response histograms for the two detectors, (a) detector 1 and (b) detector 2, with Gaussian fits used to extract the timing jitter.}
    \label{fig:timing_jitter}
\end{figure}

The timing jitter of the detection hardware is a critical parameter for temporal-filtering experiments. To characterise it, we intentionally allow a small fraction of the $\sim5$\,ps excitation laser pulses to pass through the cross-polarised microscope with the QD tuned out of resonance (using the gate voltage). The resulting instrument response function is recorded, and the timing jitter is extracted from a Gaussian fit (Fig.~\ref{fig:timing_jitter}). 
% A Gaussian fit is used rather than a convolution of the laser pulse with the detector response because the multi-photon nature of the laser pulse ensures that the first photon triggers detection, making the measurement insensitive to the pulse temporal profile. 
We define the timing jitter as the full width at half maximum (FWHM) of the Gaussian fit, obtaining values of $35.1 \pm 0.4$~ps for channel~1 and $34.6 \pm 0.4$~ps for channel~2.

\pagebreak 

\section{Temporal filtering}
\begin{figure}[h!]
    \centering
    \includegraphics{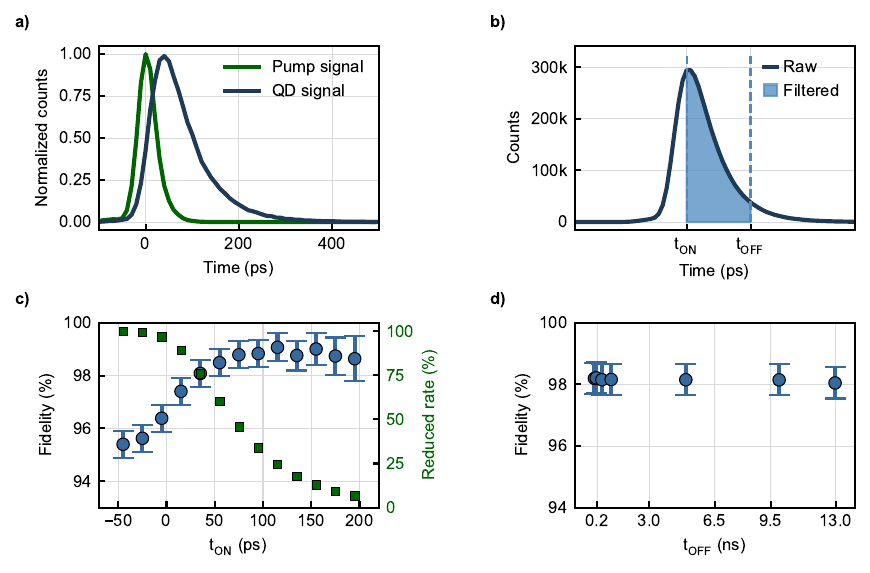}
    \caption{Temporal filtering analysis for optimising entanglement fidelity. (a) Histograms of photon detection events relative to the pump laser trigger, recorded for the pump laser alone (green) and for QD emission (blue). The maximum of the pump laser peaks defines the $t_\mathrm{ON} = 0$ reference. (b) Effect of applying the temporal filter: the full histogram of QD detection events (dark blue) and the events retained by the inclusion window (light blue shaded region) with $t_\mathrm{ON} = 35$~ps and $t_\mathrm{OFF} = 235$~ps. Only retained events are used for coincidence counting and state tomography. (c) Entanglement fidelity as a function of $t_\mathrm{ON}$ (left axis) and corresponding reduction in detection rate (right axis, dark green). (d) Entanglement fidelity as a function of $t_\mathrm{OFF}$ for $t_\mathrm{ON} = 35$~ps. 
    %For windows narrower than 100~ps, insufficient coincidences are detected for reliable tomography. 
    The plateau indicates that unwanted early-time emission is concentrated near the excitation pulse, with minimal background at later times.}
    \label{fig:temporal_filter}
\end{figure}

Temporal filtering is implemented in post-processing using raw time-tag data from the Swabian time tagger. Detection events are recorded as time-stamped photon arrivals and saved in Swabian Instruments .ttbin format. The filtering procedure applies a rectangular temporal mask to the data stream, accepting only events within a window spanning from $t_\mathrm{ON}$ to $t_\mathrm{OFF}$, where $t_\mathrm{ON}$ marks the start and $t_\mathrm{OFF}$ the stop of the inclusion window. Coincidence analysis and state tomography are then performed on the filtered dataset, i.e., only detection events falling within the [$t_\mathrm{ON}$, $t_\mathrm{OFF}$] range. As we want to exclude residual multi-photon events concentrated in time near the excitation pulse, we must first establish a reference arrival time for the excitation pulse. We do this by allowing a small fraction of the excitation laser pulse through the cross-polarised microscope and into the detection channel, with the QD gate voltage detuned from resonance. The resulting histogram defines our $t_\mathrm{ON} = 0$ reference. Figure\,\ref{fig:temporal_filter}(a) shows the excitation laser pulse relative to the QD emission.
% To establish the $t_\mathrm{ON} = 0$ reference, we record histograms of photon arrival times under two conditions: first, with the QD detuned from resonance by adjusting the bias voltage, so that only a small fraction of the excitation laser is detected (blue); and second, with QD emission enabled (green), as shown in Fig.\,\ref{fig:temporal_filter}(a). 
We define the $t_\mathrm{ON} = 0$ reference as the time bin corresponding to the peak intensity of the pump laser histogram. Typically, $t_\mathrm{OFF}$ is set such that the inclusion window spans one full repetition period, i.e., $t_\mathrm{OFF} = T_{R_L} - t_\mathrm{ON}$, where $T_{R_L} = 1/R_L \approx 13.1$~ns. This ensures that as $t_\mathrm{ON}$ increases, the inclusion window narrows accordingly to avoid including detection events from the subsequent excitation pulse. Figure~\ref{fig:temporal_filter}(b) shows the effect of applying the temporal filter: the full histogram of QD detection events is shown in dark blue, while the light blue shaded region indicates the events retained by the inclusion window with $t_\mathrm{ON} = 35$~ps and $t_\mathrm{OFF} = 235$~ps. Only the retained events are used for coincidence counting and state tomography. Figure~\ref{fig:temporal_filter}(c) presents the entanglement fidelity as a function of $t_\mathrm{ON}$, with the secondary axis showing the corresponding reduction in detection rate (dark green). As expected, increasing $t_\mathrm{ON}$ reduces the inclusion window and therefore the detection rate. Figure~\ref{fig:temporal_filter}(d) shows the fidelity as a function of $t_\mathrm{OFF}$. 
% For windows narrower than 250~ps, insufficient coincidences are detected for reliable tomography. 
The fidelity remains relatively constant as the window widens, indicating that unwanted emission is concentrated at early times near the excitation pulse.
%This ensures that as $t_\mathrm{ON}$ increases, the inclusion window narrows accordingly to avoid including detection events from the subsequent excitation pulse.

\pagebreak

\section{Density matrices}
The density matrices are reconstructed by measuring coincidence counts between detector pairs across all relevant polarisation bases. The tomographic reconstruction follows the maximum likelihood method described in Ref.~\cite{state_tomo} and is implemented in Python. The complete density matrices are shown below. Each figure displays the real part (a), imaginary part (b), and singlet fraction (c) of the reconstructed density matrix in the $\{|HH\rangle, |HV\rangle, |VH\rangle, |VV\rangle\}$ basis. The singlet fraction is obtained by applying single-qubit rotations that optimise the overlap with the target Bell state; this results in a density matrix that has only real components.

\begin{figure}[t]
    \centering
    \includegraphics{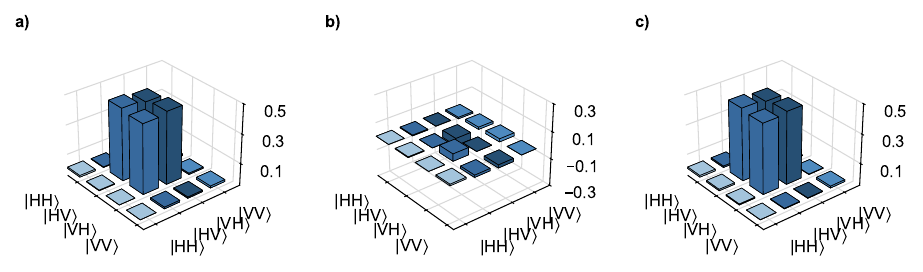}
    \caption{Quantum state tomography of the generated entangled photon pairs under standard excitation at $76.3$~MHz repetition rate with no temporal filtering. (a) Real and (b) imaginary components of the reconstructed density matrix $\rho$ in the $\{|HH\rangle, |HV\rangle, |VH\rangle, |VV\rangle\}$ basis. (c) Singlet fraction, demonstrating fidelity of $96.1 \pm 0.5\%$.}
    \label{fig:best_tomo}
\end{figure}

\begin{figure}[h]
    \centering
    \includegraphics{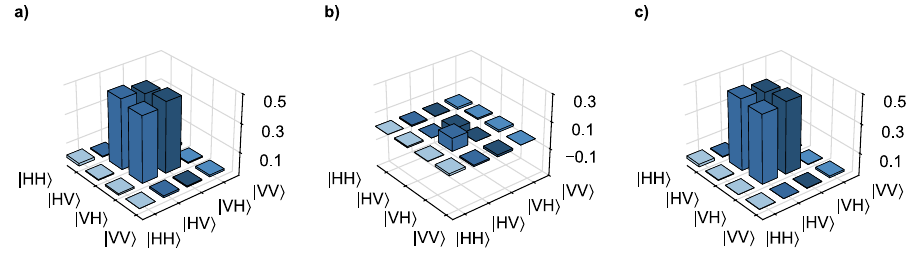}
    \caption{Quantum state tomography of the generated entangled photon pairs under high-rate excitation at an effective 2-GHz repetition rate with no temporal filtering. (a) Real and (b) imaginary components of the reconstructed density matrix $\rho$ in the $\{|HH\rangle, |HV\rangle, |VH\rangle, |VV\rangle\}$ basis. (c) Singlet fraction, demonstrating fidelity of $95.2 \pm 0.5\%$.}
    \label{fig:pulse_doubled_tomo}
\end{figure}

\begin{figure}[b]
    \centering
    \includegraphics{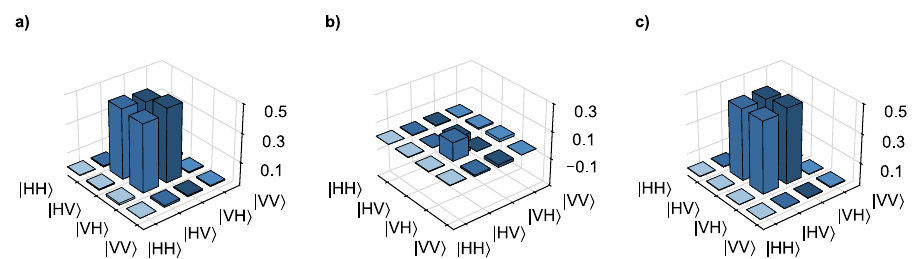}
    \caption{Quantum state tomography of the generated entangled photon pairs under standard excitation at $76.3$~MHz repetition rate after applying temporal filtering with $t_\mathrm{ON} = 35$~ps. (a) Real and (b) imaginary components of the reconstructed density matrix $\rho$ in the $\{|HH\rangle, |HV\rangle, |VH\rangle, |VV\rangle\}$ basis. (c) Singlet fraction, demonstrating fidelity of $98.1 \pm 0.5\%$.}
    \label{fig:temp_filter_tomo}
\end{figure}

Figure~\ref{fig:best_tomo} shows the reconstructed density matrix under standard excitation conditions (76.3~MHz) with no temporal filtering, achieving an entanglement fidelity of $96.1 \pm 0.5\%$. To demonstrate scalability to higher rates, Fig.~\ref{fig:pulse_doubled_tomo} presents the state tomography at an effective 2-GHz repetition rate (two pump pulses separated by $500$~ps) with no temporal filtering, maintaining a high fidelity of $95.2 \pm 0.5\%$. Finally, Fig.~\ref{fig:temp_filter_tomo} shows the density matrix obtained after applying temporal post-selection with $t_\mathrm{ON} = 35$~ps and $t_\mathrm{OFF} = 13\,020$~ps to the standard excitation data, yielding a fidelity of $98.1 \pm 0.5\%$.

\pagebreak 
\clearpage 

\section{Fitting Functions}
\subsection{Entanglement fidelity versus temporal offset}
In Figure 3(a) of the main text, we show the relationship between the entanglement fidelity of our post-selected entanglement source and the temporal offset between the two single photons interfering to create the final state. For two single photons with indistinguishability $\mathcal{I}$, the fidelity $F$ of the resulting post-selected entangled state is simply 
\begin{equation*}
F = \frac{1 + \mathcal{I}}{2}.
\end{equation*}
We are therefore interested in a formula connecting $\mathcal{I}$ and the temporal offset $\tau$. We employ the same model as \cite{brash2026}, presented here for completeness. In this case, $\mathcal{I}$ is the overlap integral between the temporal modes of the two photon wavefunctions:
\begin{equation}
    \mathcal{I}(\tau) = \mathcal{I}_0\int^{\infty}_{-\infty} \psi^*_a(t) \psi_b(t - \tau)dt,
\end{equation}
with $\psi_a(t)$ and $\psi_b(t)$ the wavefunction of the single photons at each input mode of the entangling beamsplitter and $\mathcal{I}_0$ the contribution to indistinguishability from other degrees of freedom. We assume $\psi_a(t) = \psi_b(t)$ since each photon is derived from the same quantum dot.

The intensity distribution of the quantum dot photons is modelled as the convolution of a Gaussian pulse and exponential decay:
\begin{equation}
    f(t, K) = \frac{1}{2K}\exp\big(\frac{1}{2K^2} - \frac{t}{K}\big)\text{erfc}\big(\frac{t - \frac{1}{K}}{\sqrt{2}}\big),
\end{equation}
with $K = \frac{T_1}{w_p}$, where $T_1$ is the  lifetime of the quantum dot and $w_p$ is the pulse width of the excitation laser, and erfc$(t)$ is the complimentary error function
\begin{equation*}
    \text{erfc}(x) = 1 - \text{erf}(x) = \frac{2}{\sqrt{\pi}}\int^{\infty}_x \exp(-t^2) dt.
\end{equation*}

\subsection{Entanglement fidelity versus photon purity}
To fit the data show in Figure 3(b) of the main text, we implement a simplified model of the post-selected source. The quantum dot source has the probabilities $P_0$, $P_1$, and $P_2$ of emitting zero, one, or two photons, respectively. The first photon emitted from the quantum dot in each pulse is assumed to have indistinguishability $\mathcal{I}$, and we treat the additional noise photon in the two-photon term as a completely distinguishable background with the same polarization as the primary photon. To relate $P_0$, $P_1$, and $P_2$ to $g^{(2)}(0)$, we assume $g^{(2)}(0) \ll 1$ and the mean-photon-number per pulse of the source $\mu \approx 1$. These assumptions, along with the fact that
$g^{(2)}(0) \approx  \frac{2P_2}{\mu^2}$ for small $g^{(2)}(0)$, allow us to write:
\begin{align} 
    \label{eq:g2_p2}P_2 &\approx \frac{g^{(2)}(0)}{2}, \\
    P_1 &\approx \mu - 2P_2 = 1 - g^{(2)}(0),\\
    \label{eq:g2_p0}P_0 &\approx 1 - P_1 - P_2 = \frac{g^{(2)}(0)}{2}.
\end{align}
We assume the extraction efficiency $\eta_{ex}$, coupling efficiency $\eta_{c}$, and detector efficiency $\eta_{det}$ are all symmetric for each arm of the system, such that the total probability of each photon being detected is 
\begin{equation*}
    \eta = \eta_{ex} \eta_c \eta_{det}.
\end{equation*}
For the sake of simplicity, we assume $\eta \ll 1$ and discard all terms $\mathcal{O}(\eta^3)$ in which three or more photons are detected at once.

For a given pulse of the quantum dot source, there are four possible outcomes at the detectors:
\begin{align*}
    P(\text{Two photons are detected}) &\equiv X_2 =  P_2 \eta^2\\
    P(\text{Quantum dot photon is detected}) &\equiv X_Q = P_1\eta + P_2 \eta(1-\eta)\\
    P(\text{Background photon is detected}) &\equiv X_B = P_2 \eta(1-\eta)\\
    P(\text{No photons detected}) &\equiv X_0 = P_0 + P_1 (1-\eta) + P_2 (1-\eta)^2.
\end{align*}

The post-selected entanglement source consists of two quantum dot pulses of orthogonal polarization. In the case that two quantum dot photons are detected, the corresponding state is
\begin{equation*}
    \rho_{Q} = \frac{1}{2}\begin{pmatrix}
0 & 0 & 0 & 0 \\
0 & 1 & -\mathcal{I} & 0 \\
0 & -\mathcal{I} & 1 & 0 \\
0 & 0 & 0 & 0
\end{pmatrix},
\end{equation*}
with fidelity $F_Q = \frac{1 + \mathcal{I}}{2}$. When both photons from a single pulse are detected, we measure the mixed state  $\rho_{B,0} = \frac{1}{2}(\ket{HH}\bra{HH} + \ket{VV}\bra{VV})$, since both photons will share the same polarization but are completely distinguishable. This state has no overlap with the maximally entangled state $\ket{\Psi^-} = \frac{1}{\sqrt{2}} (\ket{HV} - \ket{VH})$, and so has fidelity $F_{B,0} = 0$. Finally, by detecting either one quantum-dot photon and one background photon, or two background photons, we project onto the state $\rho_{B, \frac{1}{2}} = \frac{1}{2}(\ket{HV}\bra{HV} + \ket{VH}\bra{VH})$, with fidelity $F_{B,\frac{1}{2}} = \frac{1}{2}$. The probabilities of each of these events occurring are:
\begin{align*}
    P(\rho_{Q}) &= \frac{1}{2} X_Q X_Q,\\
    P(\rho_{B,0}) &= X_2 X_0,\\
    P(\rho_{B, \frac{1}{2}}) &= X_Q X_B + \frac{1}{2} X_B X_B.\\
\end{align*}
The final weighted fidelity is
\begin{equation*}
    F(\mathcal{I}, g^{(2)}(0)) = \frac{P(\rho_Q)\frac{1 + \mathcal{I}}{2} + P(\rho_{B,\frac{1}{2}})\frac{1}{2}}{P(\rho_Q) + P(\rho_{B,\frac{1}{2}}) + P(\rho_{B, 0})}.
\end{equation*}
For the data in Figure 3(b) of the main text, the optimal fit corresponds to indistinguishability $\mathcal{I} = 0.968$. While the fit matches the data well for small $g^{(2)}(0)$, our assumption $g^{(2)}(0) \ll 1$ leads to a worse fit as $g^{(2)}(0)$ increases.

\section{Entanglement-Swapping Model details}
To model the viability of our source in the context of quantum-networking applications, we follow the prescription of \cite{Johnson2026}. In each pulse, the quantum dot emits the single-photon state:

\begin{align*}
    \ket{\psi_{QD}} &= \sqrt{P_0}\ket{0} + \sqrt{P_1}\ket{1} + \sqrt{P_2}\ket{2},
\end{align*}
where $p(k)$ is the probability of emitting $k$ photons in the pulse, determined by the $g^{(2)}(0)$ as in Equations \ref{eq:g2_p2}-\ref{eq:g2_p0}.

This Fock-basis representation of the state accounts for the errors introduced by multi-photon emission, one of the primary noise sources for spontaneous parametric downconversion (SPDC)-based entanglement sources. After the fast switch and delay line, there are two photonic modes, each populated by $\ket{\psi_{QD}}$,
\begin{equation*}
    \ket{\psi_{DL}} = \ket{\psi_{QD}}_{H, a} \ket{\psi_{QD}}_{V,b},
\end{equation*}
where $a$ and $b$ are the two spatial input of the 50:50 beamsplitter and $H$/$V$ are the polarisations of the photons. The two-photon state is mixed on the 50:50 beamsplitter, modelled via the operator 
\begin{align*}
    B &= \sum_{m,n,k} \frac{\sqrt{(n-k)!}\sqrt{(m+k)!}}{2^{\frac{m+n}{2}}\sqrt{n!}\sqrt{m!}} A_k |m-k\rangle_{X,1} |n+k \rangle_{X,2} \langle m|_{X,1} \langle n|_{X,2},\\
    \nonumber A_k &= \sum_l \binom{m}{k+l}\binom{n}{l} i^{k+2j}.
\end{align*}
The final state 
\begin{equation*}
    \ket{\psi} = B(\ket{\psi_{DL}})
\end{equation*}
is the output of the entanglement source, with post-selection via state-projection to remove the vacuum, recovering a maximally entangled Bell state for $g^{(2)}(0) = 0$. This state is then used as input to the entanglement-swapping model described in \cite{Johnson2026} to produce the plot shown in Fig.~5 of the main paper.